\documentclass[prl,twocolumn,superscriptaddress,nopacs]{revtex4}
\makeatletter
\def\@dotsep{4.5}
\makeatother

\usepackage{graphics}

\newcommand{\kB}{k_{\mathrm{B}}}

\newcommand{\ffrac}[2]{\frac{\displaystyle #1}{\displaystyle #2}}

\newcommand{\fig}[4]
{
\begin{figure}[#4]
\resizebox{\columnwidth}{!}{\includegraphics{#1}}
\caption{\label{#3}#2}          
\end{figure}
}

\newcommand{\figs}[4]
{
\begin{figure}[#4]
\resizebox{\columnwidth}{!}{\includegraphics{#1}}
\end{figure}
}

\begin{document}

\title{De-coupling of Exchange and Persistence Times in Atomistic Models of Glass Formers}

\author{Lester O. Hedges}

\affiliation{School of Physics and Astronomy, University of Nottingham,
Nottingham, NG7 2RD, UK}

\author{Lutz Maibaum}

\affiliation{Department of Chemistry, University of California,
Berkeley, CA 94720-1460}

\author{David Chandler}

\affiliation{Department of Chemistry, University of California,
Berkeley, CA 94720-1460}

\author{Juan P. Garrahan}

\affiliation{School of Physics and Astronomy, University of Nottingham,
Nottingham, NG7 2RD, UK}

\begin{abstract}
With molecular dynamics simulations of a fluid mixture of classical particles interacting with pair-wise additive Weeks-Chandler-Andersen potentials, we consider the time series of particle displacements and thereby determine distributions for local persistence times and local exchange times.   These  basic characterizations of glassy dynamics are studied over a range of super-cooled conditions and shown to have behaviors, most notably de-coupling, similar to those found in kinetically constrained lattice models of structural glasses.  Implications are noted.
\end{abstract}


\maketitle

Facilitated dynamics, as encoded in various kinetically constrained lattice models (KCMs),\cite{FA, Ritort-Sollich} implies a general picture of structural glasses in which never-ending excitation lines coalesce, branch and percolate throughout space-time.\cite{JPG&DC-PRL02, JPG&DC-PNAS03}  Particles cannot move except when intersected by these excitation lines.\cite{Jung1, Jung2, Berthier1}  As such, principal signatures of glassy dynamics can be viewed as manifestations of the distribution of times for which a particle must wait to first encounter an excitation, and the distribution of times between subsequent encounters.  These are what we have called the distributions of ``persistence'' times and ``exchange'' times, respectively.\cite{Jung1, Jung2, Berthier1}  Definitions and behaviors of excitation lines, persistence and exchange are most obvious in idealized lattice models.  Nevertheless, these concepts have proved useful for interpreting intermittent and heterogeneous dynamics and transport de-coupling\cite {SE, DH} in continuous force model systems and real systems,\cite{Jung1, Berthier1, chi4, Berthier2} suggesting that these concepts are not limited to KCMs.  Here, we show that indeed, exchange and persistence are well defined in terms of classical trajectories of the sort computed from molecular dynamics or observed with microscopy,\cite{Colloids} and for a specific continuous force model, the distributions for exchange and persistence times behave similarly to those we have previously gleaned from KCMs.\cite{Jung2,Albert}

Figure 1 illustrates our main result.  It  shows the distributions of persistence and exchange times for particle displacement events obtained from molecular dynamics simulations of an atomistic model.  A local dynamical event or excitation is defined as a displacement beyond a specified cutoff length (see below). As the liquid becomes increasingly super-cooled the two distributions de-couple, and the typical persistence time becomes much larger than the typical exchange time.    This de-coupling occurs because excitations or mobility and their associated excitation lines become relatively sparse at low temperatures. As such, for low enough temperatures, the time extent of space-time regions devoid of excitation lines (regions  that dominate persistence processes) is typically very long compared to that for regions bridging the width of excitation lines (regions that dominate exchange processes).  Thus, the origin of this de-coupling is the same as that of dynamic heterogeneity.\cite{JPG&DC-PRL02, JPG&DC-PNAS03, Jung1}

This de-coupling is reflected, for example, in observed breakdowns \cite{SE} in mean-field transport relations like the Stokes-Einstein inverse proportionality between diffusion constant, $D$, and structural relaxation time, $\tau_\alpha$.\cite{chi4-decoupling}  In particular, when de-coupling occurs, diffusion is much faster than would be predicted from $D\propto1/\tau_\alpha$ because diffusion, being an exchange process, has $1/D$ proportional to the first moment of the exchange-time distribution while $\tau_\alpha$ is the first moment of persistence-time distribution.\cite{Jung1}

\fig{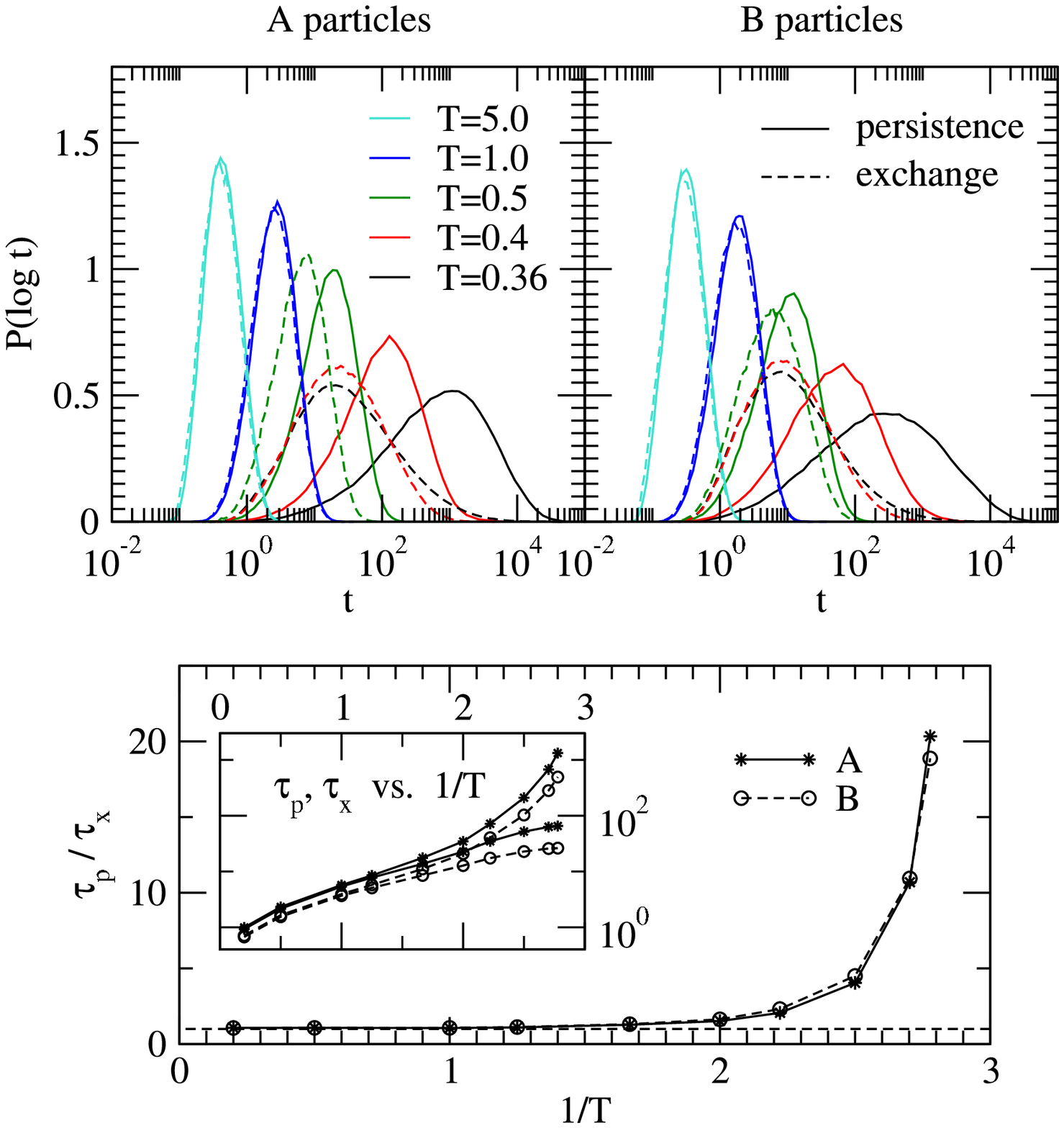}{De-coupling of exchange and persistence times in the WCA mixture. A local event is defined as a particle moving a distance larger than $d$.  We show results for $d=0.5$. (Top) Distributions of exchange times and of persistence times for particle species $A$ and $B$ for various temperatures $T$.   (Bottom) Ratio of the average persistence time, $\tau_{\rm p} \equiv \langle t_{\rm p} \rangle$, to the average exchange time, $\tau_{\rm x} \equiv \langle t_{\rm x} \rangle$, as a function of $T$.  The inset shows $\tau_{\rm p}$ and $\tau_{\rm x}$ for both species as a function of $1/T$.}{fig1}{hb}

The model \cite{Lutz} we study is a variation of the binary Lennard--Jones mixture of Ref.\ \onlinecite{Wahnstroem91}, which has been extensively studied as a model super-cooled liquid (see e.g. Ref.\ \onlinecite{Lacevic}).  We modify this system by removing the attractive part of the Lennard--Jones interaction by adopting the Weeks-Chandler-Andersen (WCA) separation of the pair-potential.\cite{WCA} We consider a mixture of two particle species $A$ and $B$ in a cubic simulation box of side length $L$ and volume $V=L^3$. The potential energy is the sum of the pairwise interactions between two particles of species $\mu, \nu \in \left\{A,B\right\}$,
\begin{equation}
\label{eq:WCAinteractions}
V_{\mu\nu} (r) =  4 \varepsilon_{\mu\nu} \left[ \left(\ffrac{\sigma_{\mu\nu}}{r}\right)^{12} - \left(\ffrac{\sigma_{\mu\nu}}{r}\right)^{6} + \ffrac{1}{4}\right]
\end{equation} 
if their separation $r$ is less than $2^{1/6} \sigma_{\mu\nu}$, and $V_{\mu\nu} (r) = 0$ otherwise. Following Ref.\ \onlinecite{Wahnstroem91} we choose $\sigma_{AA} = 1, \sigma_{BB} = 5/6, \sigma_{AB} = (\sigma_{AA} + \sigma_{BB})/2$ and $\varepsilon_{AA} = \varepsilon_{BB} = \varepsilon_{AB} = 1$. The particle masses are $m_A = 2$ and $m_B = 1$. Lengths, times and temperatures are reported in units of $\sigma_{AA}$, $(m_B \sigma_{AA}^2 / \varepsilon_{AA})^{1/2}$ and $\varepsilon_{AA} / \kB$, respectively. The number of particles of species $\mu$ is $N_\mu$, and the corresponding mole fraction is $N_\mu / N = N_\mu / (N_A + N_B)$.

Due to the shortness of the interaction range, $2^{1/6} \sigma \approx 1.12 \sigma$, each particle interacts with only its nearest neighbors, significantly reducing the computational overhead of calculating forces.  We find that this feature makes the WCA mixture up to one order of magnitude faster to simulate than the corresponding Lennard--Jones mixture.\cite{Lutz}  It helps us study a reasonably large system at significantly super-cooled conditions.

\fig{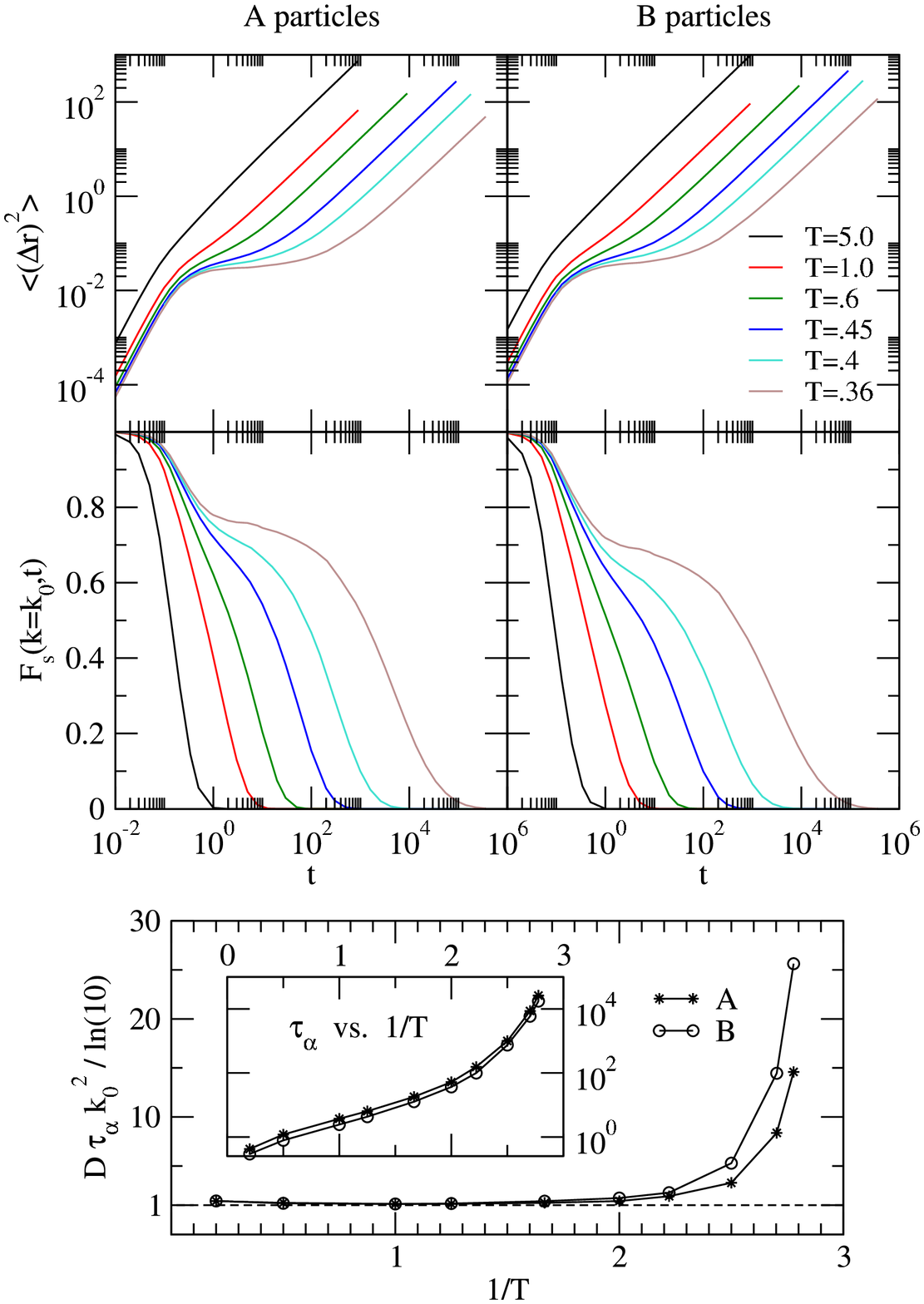}{(Top) Mean squared displacement $\langle (\Delta r)^2 \rangle = \langle |\Delta{\bf r}(t)|^2 \rangle$ for both particle species $A$ and $B$ as a function of time $t$ for various temperatures $T$.  (Center) Incoherent scattering function $F_s(k,t) = \langle e^{i {\bf k} \cdot \Delta {\bf r}(t)} \rangle$, evaluated at the wavevector $k=k_0$ of the first peak in the partial structure factors, for both species at the same temperatures as above.  (Bottom) Violation of the Stokes-Einstein relation (dashed line) for decreasing temperatures.  The inset shows the corresponding relaxation times $\tau_\alpha$, defined by the relation $F_s (k_0, \tau_\alpha) = 0.1$, for each species as a function of inverse temperature $1/T$.}{fig2}{t}

Figure 2 demonstrates that the WCA mixture has the standard phenomenology associated with glass formation.  The figure presents results from two-point time-correlation functions for varying temperatures ranging from the normal liquid regime to the super-cooled regime.\cite{MD-details}  Figure\ 2(top) and Fig.\ 2(center) show the mean-squared displacement $\langle |\Delta{\bf r}(t)|^2 \rangle$ and the self-intermediate scattering function $F_s(k,t) \equiv \langle e^{i {\bf k} \cdot \Delta {\bf r}(t)} \rangle$ for $k$ at the peak of the static structure factor $k=k_0$, respectively, as a function of time.  The curves display the characteristic low temperature features of increasing slowing down, plateaus and stretching observed in simulations of similar systems.\cite{Wahnstroem91,Simus}
Static density correlations are stationary over the 
range of temperatures considered, which suggests nothing of thermodynamic significance is 
happening throughout the simulation.  The density correlation functions decay to their uncorrelated 
values at lengths much smaller than the simulation box length, and its structure factor shows no 
anomalies at small wave vectors.  These facts indicate that the system remains a liquid mixture
throughout the simulation and does not evolve towards a 
crystallized or phase-separated state. Dynamical quantities show typical liquid-like 
behavior when measured over long enough time scales.  For instance, the small wave vector incoherent scattering function is fully consistent with diffusive relaxation. This fact also supports the conclusion that the system is not evolving towards a crystallized or phase-separated state.\cite{Lutz}

Figure\ 2(bottom) shows the breakdown of the Stokes-Einstein relation at low temperatures.  At the lowest temperature we have studied, the de-coupling between the diffusion constant and the structural relaxation time is about an order of magnitude.  This again is similar to what is seen in similar systems.\cite{SE}  A detailed study of dynamic heterogeneity in the WCA mixture is given in Ref.\ \onlinecite{Lutz} and is left to a future publication.

To obtain the distributions shown in Fig.\ 1, we monitor the time series of events for each particle in the system, defining an event as a particle is displaced beyond a cutoff length $d$.  Consider particle $i$.  At the initial time $t=0$, when we start the observation, its position is ${\bf r}_i(0)$.  The first event time for that particle, $t_1$, is the {\em first} time that particle $i$ has moved far enough so that $| {\bf r}_i(t_1) - {\bf r}_i(0)| = d$.  A second event occurs for that particle at time $t=t_1+t_2$, when particle $i$ manages to move again a distance $d$, this time from its position at $t_1$; i.e., $| {\bf r}_i(t_1+t_2) - {\bf r}_i(t_1)| = d$.  A third event takes place after a further wait $t_3$, and so on.  

This convention establishes a set of waiting times between events for the $i$-th particle, $\{ t_1, t_2, t_3, \ldots \}$.  Notice, however, that $t_1$ has an important physical difference from $t_2, t_3, \ldots$.  The time $t_1$ is the time for the first event to take place {\em without condition} on when the previous event occurred.  The times $t_2, t_3, \ldots$ are times {\em between} events.  As noted, we call the time $t_1$ a {\em persistence} time, and the times $t_2, t_3, \ldots$ {\em exchange} times.\cite{Renewal}

Figure\ 1(top) shows the distributions of exchange and persistence times for the two kinds of particles in the WCA mixture, for temperatures ranging from the normal liquid regime to the super-cooled one.  We choose the cutoff length $d$ to be comparable to the particle size, so as to probe particle motion relevant to structural relaxation and diffusion.  As argued in previous work,\cite{JPG&DC-PRL02,JPG&DC-PNAS03} dynamic facilitation features emerge only after a suitable coarse-graining of short scale motion.  The length $d$ plays the role of a coarse-graining length.  The distributions of Fig.\ 1 are for $d=0.5$.  We have obtained similar results for $d$ in the range $d=0.5$--$1.0$.\cite{Allegrini}
  
At high temperatures, the exchange and persistence time distributions coincide.  This is the situation when the distribution of exchange times is exponential.\cite{Jung2}  As the temperature is decreased the two distributions become distinct, that for persistence moving towards longer times.   Fig.\ 1(bottom) shows the ratio $\tau_{\rm p} / \tau_{\rm x}$, where 
$\tau_{\rm p} \equiv \langle t_{\rm p} \rangle$ is the average persistence time and 
$\tau_{\rm x} \equiv \langle t_{\rm x} \rangle$ is the average exchange time.   At low temperatures the average persistence time becomes much larger than the exchange time.  This de-coupling mirrors that of the breakdown of the Stokes-Einstein relation of Fig.\ 2.  

The de-coupling of exchange and persistence times in the WCA mixture is similar to that observed in KCMs.\cite{Jung2,Albert}  These distributions reflecting the time series of individual particle displacement events are non-Poissonian.  Event times cluster with periods of quiescence followed by periods of high activity.  This clustering of event times is one way to view dynamic heterogeneity, or more precisely, a coexistence in space-time between active and inactive dynamical phases.\cite{DF}   Phase separation in trajectory space is a distinguishing prediction of dynamic facilitation.\cite{DF} De-coupling of persistence and exchange processes is a consequence of this phase separation.  

By describing dynamics in terms of exchange and persistence, one is able to picture particle motion as a continuous-time random walk.\cite{Montroll}  An individual particle makes random walk steps at random times drawn from the persistence and exchange time distributions.\cite{Jung1}  Along with accounting for the breakdown of Stokes-Einstein relations,\cite{Jung1} this simple picture also accounts for the non-Fickian to Fickian crossover\cite{Berthier1} and the shape\cite{Stariolo} of the van Hove correlation function in a wide range of systems.\cite{Berthier2}

\bigskip
In carrying out this work JPG was supported by EPSRC grant GR/S54074/01, LM was supported by DOE grant under Contract No. DE-AC02-05CH11231, and DC was initially supported by NSF grant No. CHE-0543158 and currently supported by ONR grant N00014-07-1-0689.  Calculations were made possible through access to the NERSC super-computer facility at Lawrence Berkeley National Laboratory.


\begin{thebibliography}{99}
\bibitem{chi4-decoupling} The decoupling of the different relevant relaxation rates is also evident from the behaviour of four-point dynamic susceptibilities at different lengthscales, see for example Refs.\ \onlinecite{chi4,Glotzer,Berthier-others}.
\bibitem{MD-details} Molecular dynamics simulations were performed using a microcanonical Verlet algorithm for Fig.\ 1 and LAMMPS\cite{LAMMPS} for Fig.\ 2, at temperatures $T=5.0$ to $0.36$ along a constant density path.  The particle
numbers and box length, $N_A = N_B = 500$ and $L=9.172$ for Fig.\ 1, and $N_A = N_B = 4000$ and $L = 18.344$ for Fig.\ 2, were fixed in each
run.  The system was prepared at each thermodynamic state point by coupling to a
heat bath for time $t_{\mathrm{cool}}$ using velocity rescaling or velocity reassigning every 50 time steps. No drift or aging was observed over the equilibration time $t_{\mathrm{equil.}}$ in thermodynamic or dynamic observables before performing production runs. Both $t_{\mathrm{cool}}$ and $t_{\mathrm{equil.}}$ were significantly longer than the structural relaxation time.
\bibitem{Renewal}
In the context of renewal processes\cite{Grimmett} exchange and persistence times are called total and excess lifetimes, respectively.  If events occur over a very short time scale as compared to waiting times the distributions of persistence and exchange times are related by $p(t_{\rm p}) = \langle t_{\rm x} \rangle^{-1} \int_{t_{\rm p}}^{\infty} p(t_{\rm x}) dt_{\rm x}$.\cite{Grimmett,Jung2}  In our case, where an event may be built up over time (a displacement of distance $d$ does not occur instantaneously), this relation works only as an approximation.  Nevertheless, application of this formula gives reasonable estimates of the behaviour.
\end{thebibliography}
\end{document}